\documentclass[11pt]{llncs}
\pdfoutput=1
\usepackage{enumerate}
	\usepackage[pdftex]{graphicx}

	\DeclareGraphicsExtensions{.pdf, .jpg, .tif}
\begin{document}

\title{Distributed Relay Protocol for Probabilistic Information-Theoretic Security in a Randomly-Compromised Network}

\author{Travis R. Beals\inst{1} \and Barry C. Sanders\inst{2}}
\institute{Department of Physics, University of California, Berkeley, California 94720, USA \and Institute for Quantum Information Science, University of Calgary, Alberta T2N 1N4, Canada}

%\date{\today}

\maketitle

\begin{abstract}
We introduce a simple, practical approach with probabilistic in\-for\-ma\-tion-theoretic security to mitigate one of quantum key distribution's
major limitations: the short maximum transmission distance ($\sim200$ km) possible with present day technology. Our scheme uses classical secret sharing techniques to allow secure transmission over long distances through a network containing randomly-distributed compromised nodes. The protocol provides arbitrarily high confidence in the security of the protocol, and modest scaling of resource costs with improvement of the security parameter. Although some types of failure are undetectable, users can take preemptive measures to make the probability of such failures arbitrarily small.
\\

\end{abstract}

\section{Introduction\label{sec:introduction}}

Public key cryptography is a critical component of many widely-used cryptosystems, and forms the basis for much of our ecommerce transaction security infrastructure. Unfortunately, the most common public key schemes are known to be insecure against quantum computers. In 1994, Peter Shor developed a quantum algorithm for efficient factorization and discrete logarithms~\cite{shor:factor2}; the (supposed) hardness of these two problems formed the basis for RSA and DSA, respectively. Sufficiently powerful quantum computers do not yet exist, but the possibility of their existence in the future already poses problems for those with significant forward security requirements.

A more secure replacement for public key cryptography is needed. Ideally, this replacement would offer information-theoretic security, and would possess most or all of the favorable qualities of public key cryptography. At present, no complete replacement exists, but quantum key distribution (QKD)---in conjunction with one-time pad (OTP) or other symmetric ciphers---appears promising.

QKD---first developed by Bennett and Brassard~\cite{bennett:BB84}---is a key distribution scheme that relies upon the uncertainty principle of quantum mechanics to guarantee that any eavesdropping attempts will be detected. In a typical QKD setup, individual photons are sent through optical fiber or through free space from the sender to the receiver. The receiver performs measurements on the photons, and sender and receiver communicate via an authenticated (but not necessarily private) classical channel.

Optical attenuation of these single photon pulses limits the maximum transmission distance for a single QKD link to about 200 km over fiber with present technology~\cite{takesue:qkd}, and significantly less through air. Unlike optically-encoded classical information, the ``signal strength'' of these photons cannot be amplified using a conventional optical amplifier; the No Cloning Theorem~\cite{wootters:cloning} prohibits this. We refer to this challenge as the \emph{relay problem}.

Two classes of quantum repeaters have been proposed to resolve the distance limitations of QKD. The first makes use of quantum error correction to detect and rectify errors in specially-encoded pulses. Unfortunately, the extremely low error thresholds for such schemes ($\sim 10^{-4}$) make this impractical for use in a realistic quantum repeater. The second class of quantum repeaters uses entanglement swapping and distillation~\cite{briegel:5932,duan:6862} to establish entanglement between the endpoints of a chain of quantum repeaters, which can then be used for QKD~\cite{ekert:661}. This method is much more tolerant of errors, and offers resource costs that scale only polynomially with the number of repeaters (i.e., polynomially with distance). However, such repeaters do have one major drawback: they require quantum memories with long decoherence times~\cite{duan:6862}.

In order to be useful for practical operation, a quantum repeater must possess a quantum memory that meets the following three requirements:
\begin{enumerate}
\item Long coherence times: at a minimum, coherence times must be comparable to the transit distance for the entire repeater chain (e.g., $\sim 10\ \mathrm{ms}$ for a trans-Atlantic link).
\item High storage density: the bandwidth for a quantum repeater is limited by the ratio of its quantum memory capacity to the transit time for the entire repeater chain~\cite{simon:190503}.
\item Robustness in extreme environments: practical quantum repeaters must be able to operate in the range of environments to which telecom equipment is exposed (e.g., on the ocean floor, in the case of a trans-oceanic link).
\end{enumerate}
These requirements are so demanding that it is possible that practical quantum repeaters will not be widely available until after large-scale quantum computers have been built---in other words, not until too late.

The distance limitations of QKD and the issues involved in developing practical quantum repeaters make it challenging to build secure QKD networks that span a large geographic area. The na\"{\i}ve solution of classical repeaters leads to exponentially decaying security with transmission distance if each repeater has some independent probability of being compromised. If large QKD networks are to be built in the near future (i.e., without quantum repeaters), an alternative method of addressing the single-hop distance limitation must be found. We refer to this as the \emph{relay problem}.

Given an adversary that controls a randomly-determined subset of nodes in the network, we have developed a solution to the relay problem that involves encoding encryption keys into multiple pieces using a secret sharing protocol~\cite{shamir:secret,blakley:313}. These shares are transmitted via multiple multi-hop paths through a QKD network, from origin to destination. Through the use of a distributed re-randomization protocol at each intermediate stage, privacy is maintained even if the attacker controls a large, randomly-selected subset of all the nodes. 

We note that authenticated QKD is information-theoretic secure~\cite{renner:012332}, as is OTP; in combination, these two cryptographic primitives provide information-theoretic security on the level of an individual link. Our protocol makes use of many such links as part of a network that provides information-theoretic security with very high probability. In particular, with some very small probability $\delta$, the protocol fails in such a way as to allow a sufficiently powerful adversary to perform undetected man-in-the-middle (MITM) attacks. The failure probability $\delta$ can be made arbitrarily small by modest increases in resource usage. In all other cases, the network is secure. We describe the level of security of our protocol as \emph{probabilistic information-theoretic}.

In analyzing our protocol, we consider a network composed of a chain of ``cities'', where each city contains several parties, all of whom are linked to all the other parties in that city. We assume intracity bandwidth is cheap, whereas intercity bandwidth is expensive; intercity bandwidth usage is the main resource considered in our scaling analysis. For the sake of simplicity, we consider communication between two parties (Alice and Bob) who are assumed to be at either end of the chain of cities. A similar analysis would apply to communication between parties at any intermediate points in the network.

\section{Adversary and Network Model}

It is convenient to model networks with properties similar to those described above by using undirected graphs, where each vertex represents a node or party participating in the network, and each edge represents a secure authenticated private channel. Such a channel could be generated by using QKD in conjunction with a shared secret key for authentication, or by any other means providing information-theoretic security.

We describe below an adversary and network model similar in some ways to one we proposed earlier\footnote{Pre-print available at www.arXiv.org as arXiv:0803.2717} 
in the context of a protocol for authenticating mutual strangers in a very large QKD network, which we referred to as the \emph{stranger authentication protocol}. In that protocol, edges represented shared secret keys, whereas here they represent physical QKD links. Network structure in the previous model was assumed to be random (possibly with a power law distribution, as is common in social networks), whereas here the network has a specific topology dictated by geographic constraints, the distance limitations of QKD, and the requirements of the protocol.

\subsection{Adversarial Capabilities and Limitations\label{sec:ad_cap}}
We call the following adversary model the \emph{sneaky supercomputer}:
\begin{enumerate}[(i)]
\item \label{it:adcap1}The adversary is computationally unbounded.
\item \label{it:adcap2}The adversary can listen to, intercept, and alter any message on any public channel.
\item \label{it:adcap3}The adversary can compromise a randomly-selected subset of the nodes in the network. Compromised nodes are assumed to be under the complete control of the adversary. The total fraction of compromised nodes is limited to $(1-t)$ or less.
\end{enumerate}

Such an adversary is very powerful, and can successfully perform MITM attacks against public key cryptosystems (using the first capability) and against unauthenticated QKD (using the second capability), but not against a QKD link between two uncompromised nodes that share a secret key for authentication (since quantum mechanics allows the eavesdropping to be detected) \cite{renner:012332}. The adversary can always perform denial-of-service (DOS) attacks by simply destroying all transmitted information; since DOS attacks cannot be prevented in this adversarial scenario, we concern ourselves primarily with security against MITM attacks. Later, we will briefly consider variants of this adversarial model and limited DOS attacks.

The third capability in this adversarial model---the adversary's control of a random subset of nodes---simulates a network in which exploitable vulnerabilities are present on some nodes but not others. As a first approximation to modeling a real-world network, it is reasonable to assume the vulnerable nodes are randomly distributed throughout the network.

An essentially equivalent adversarial model is achieved if we replace the third capability as follows: suppose the adversary can attempt to compromise any node, but a compromise attempt succeeds only with probability $(1-t)$, and the adversary can make no more than one attempt per node. In the worst case where the adversary attempts to compromise all nodes, the adversary will control a random subset of all nodes, with the fraction of compromised nodes being roughly $(1-t)$.

\subsection{The Network}
For the relay problem, let us represent the network as a graph~$G$, with~$V(G)$ being the set of vertices (nodes participating in the network) and $E(G)$ being the set of edges (secure authenticated channels, e.g. QKD links between parties who share secret keys for authentication). $N = |V(G)|$ is the number of vertices (nodes). $V_d$ is the set of compromised nodes, which are assumed to be under the adversary's control; $|V_d| \leq N (1-t)$. Furthermore, let us assume that the network has the following structure: nodes are grouped into $m$ clusters---completely connected sub-graphs containing $n$ nodes each. There are thus $N=mn$ nodes in the network. We label the nodes as $v_{i,j}$, $i\in \left\{1,\dots,n\right\}$, $j\in \left\{1,\dots,m\right\}$. Each node is connected to one node in the immediately preceding cluster and one node in the cluster immediately following it. 

More formally, let $E_\ell(G) \equiv \{(v_{i,j},v_{i,j+1}) : v_{i,j}, v_{i,j+1}  \in V(G)\}$ and $E_\sigma(G) \equiv \{(v_{i,j},v_{k,j}) : v_{i,j}, v_{k,j} \in V(G)\}$. Then, $E(G) \equiv E_\ell(G) \cup E_\sigma(G)$.

This network structure models a chain of $m$ cities (a term which we use interchangeably with ``cluster''), each containing $n$ nodes. The cities are spaced such that the physical distance between cities allows QKD links only between adjacent cities. To realistically model the costs of communication bandwidth, we assume that use of long distance links (i.e., those represented by $E_\ell(G)$) is expensive, whereas intracity links (i.e., $E_\sigma(G)$) are cheap.

Next, we consider two additional nodes---a sender and a receiver. The sender (hereafter referred to as Alice or simply $A$) has direct links to all the nodes in city 1, while the receiver (Bob, or $B$) has a link to all nodes in city $m$. We assume Alice and Bob to be uncompromised. An example is shown in Fig. \ref{fig:relay_graph}.

\section{The Relay Protocol\label{sec:relay}}
In the relay problem, Alice wishes to communicate with Bob over a distance longer than that possible with a single QKD link, with quantum repeaters being unavailable. As described above, Alice and Bob are separated by $m$ ``cities'', each containing $n$ participating nodes. (In the case where different cities contain different numbers of participating nodes, we obtain a lower bound on security by taking $n$ to be the minimum over all cities.) 

\begin{figure} \centering
\includegraphics[width=3.25 in,  keepaspectratio=true]{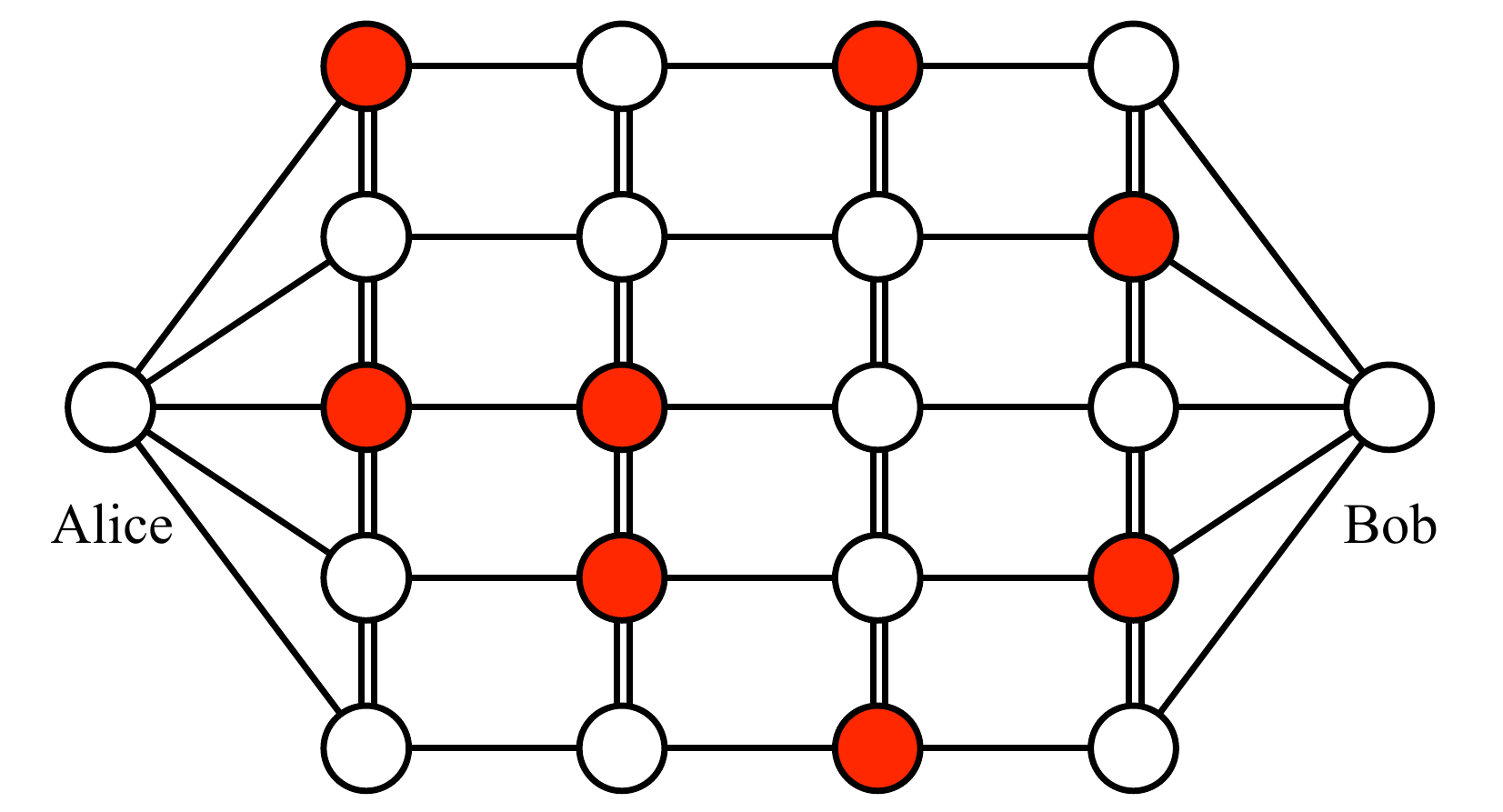}
\caption{\label{fig:relay_graph} White vertices represent honest parties, whereas shaded vertices represent dishonest parties. Double vertical lines represent secure communication links between all joined vertices (i.e., all parties within a given city can communicate securely). In the graph shown above, $40\%$ of the parties in cities between Alice and Bob are dishonest, but Alice and Bob can still communicate securely using the method described in Sec. \ref{sec:relay} and Fig. \ref{fig:protocol}.}
\end{figure}

To achieve both good security and low intercity bandwidth usage, we can employ a basic secret sharing scheme with a distributed re-randomization of the shares \cite{ben-or:distributed} performed by the parties in each city. This re-randomization procedure is similar to that used in the mobile adversary proactive secret sharing scheme \cite{ostrovsky:112605,herzberg:339}. Note that in the following protocol description, the second subscript labels the city, while the first subscript refers to the particular party within a city.

\begin{figure} \centering
 \includegraphics[width=3.50 in,  keepaspectratio=true]{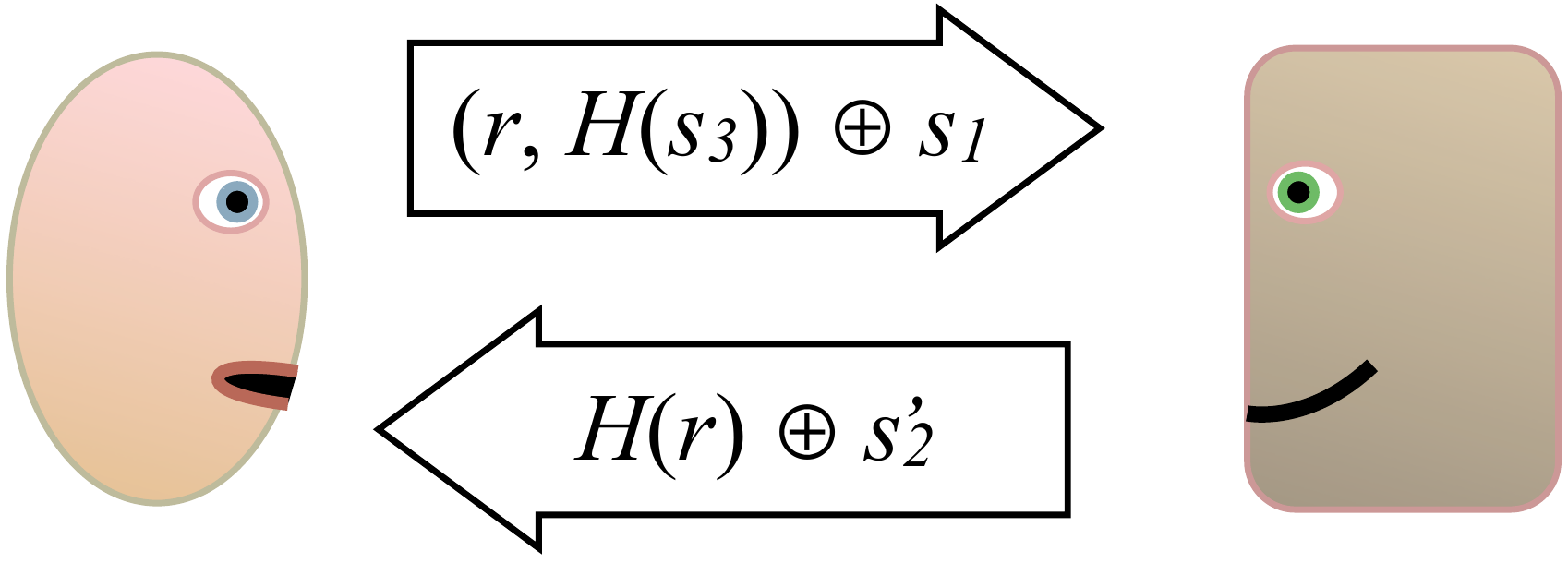}
 \caption[]{\label{fig:protocol} Alice and Bob perform a verification sub-protocol to check that their respective secret keys, $s=(s_1,s_2,s_3)$ and $s^\prime=(s_1^\prime,s_2^\prime,s_3^\prime)$, are in fact the same.  Alice generates a random number $r$, concatenates it with the hash $H[s_3]$ of $s_3$, XORs this with $s_1$, and sends the result to Bob. Bob decodes with $s_1^\prime$, verifies that $H[s_3] = H[s_3^\prime]$, then sends back to Alice the result of bit-wise XORing the hash of $r$, $H[r]$, with $s_2^\prime$. Finally, Alice decodes with $s_2$ and checks to see that the value Bob has computed for $H[r]$ is correct. Alice and Bob now know $s_3 = s_3^\prime$ and can store $s_3$ for future use. Note that with this protocol, the adversary can fool Alice and Bob into accepting $s \neq s^\prime$ with 100 \% probability if the adversary knows $s$ and $s^\prime$. }
 \end{figure}

\begin{enumerate}[(i)]
  \item Alice generates $n$ random strings $r_{i,0}, i\in\{1,\ldots,n\}$ of length $\ell$, $r \in \{0,1\}^\ell$. $\ell$ is chosen as described in Sec. \ref{sec:verify_protocol}.
  \item Alice transmits the strings to the corresponding parties in the first city: $v_{i,1}$ receives $r_{i,0}$.
  \item \label{it:party_rec}When a party $v_{i,j}$ receives a string $r_{i,j-1}$, it generates $n-1$ random strings $q_{i,j}^{(k)}, k\neq i$ of length $\ell$, and transmits each string $q_{i,j}^{(k)}$ to party $v_{k,j}$ (i.e., transmission along the vertical double lines shown in Fig. \ref{fig:relay_graph}).
  \item \label{it:party_gen}Each party $v_{i,j}$ generates a string $r_{i,j}$ as follows: 
  \[r_{i,j} \equiv r_{i,j-1} \oplus \left(\bigoplus_{k,k\neq i} q_{i,j}^{(k)} \right) \oplus \left( \bigoplus_{k,k\neq i} q_{k,j}^{(i)} \right),\]
  where the symbols ($\oplus$ and $\bigoplus$) are both understood to mean bitwise XOR. Note that the string $r_{i,j-1}$ is received from a party in the previous city, the strings $q_{i,j}^{(k)}$ are generated by the party $v_{i,j}$, and the strings $q_{k,j}^{(i)}$ are generated by other parties in the same city as $v_{i,j}$. The string $r_{i,j}$ is then transmitted to party $v_{i,j+1}$ (i.e., transmission along the horizontal lines shown in Fig. \ref{fig:relay_graph}).
  \item Steps (\ref{it:party_rec}) and (\ref{it:party_gen}) are repeated until the strings reach the parties in city $m$. All the parties $v_{i,m}$ in city $m$ forward the strings they receive to Bob.
  \item Alice constructs $s \equiv \prod_{i} r_{i,0}$ and Bob constructs $s^\prime \equiv \prod_{i} r_{i,j-1}$.
  \item Alice and Bob use the protocol summarized in Fig. \ref{fig:protocol} and described in detail in Section \ref{sec:verify_protocol} to determine if $s=s^\prime$. If so, they are left with a portion of $s$ (identified as $s_3$), which is their shared secret key. If $s \neq s^\prime$, Alice and Bob discard $s$ and $s^\prime$ and repeat the protocol.
\end{enumerate}

\subsection{Key Verification \label{sec:verify_protocol}}
In the last step of the protocol described above, Alice and Bob must verify that their respective keys, $s$ and $s^\prime$, are the same and have not been tampered with. We note that there are many ways\footnote{See for example pp. 13--14 of the SECOQC technical report D-SEC-48, by L. Salvail \cite{salvail:qkd}.} to accomplish this; we present one possible method here (summarized in Fig. \ref{fig:protocol}) for definiteness, but make no claims as to its efficiency.

We consider Alice's key $s$ to be composed of three substrings, $s_1$, $s_2$, and $s_3$, with lengths $\ell_1$, $\ell_2$, and $\ell_3$, respectively (typically, $\ell_3 \gg \ell_1,\ell_2$). Bob's key $s^\prime$ is similarly divided into $s_1^\prime$, $s_2^\prime$, and $s_3^\prime$. If Alice and Bob successfully verify that $s_3^\prime = s_3$, they can use $s_3$ as a shared secret key for OTP encryption or other cryptographic purposes.

The verification is accomplished as follows:
\begin{enumerate}[(i)]
\item Alice generates a random nonce $r$, and computes the hash $H[s_3]$ of $s3$. She then sends $(r,H[s_3]) \oplus s_1$ to Bob.
\item Bob receives the message from Alice, decrypts by XORing with $s_1^\prime$, and verifies that the received value of $H[s_3]$ matches $H[s_3^\prime]$. If so, he accepts the key, and sends Alice the message $H[r] \oplus s_2^\prime$. If not, Bob aborts. 
\item Alice decrypts Bob's message by XORing with $s_2$, and verifies that the received value of $H[r]$ is correct. If so, Alice accepts the key, and verification is successful. If not, Alice aborts.
\end{enumerate}

We now outline a proof of the security of this verification process, and discuss requirements for the hash function $H$. We begin with the assumption that Eve does not know $s$ or $s^\prime$; if she does, the relay protocol has failed, and Eve can perform MITM attacks without detection (conditions under which the relay protocol can fail are analyzed in Sec. \ref{sec:security}). Our goal is to show that Alice and Bob will with very high probability detect any attempt by Eve to introduce errors in $s_3^\prime$ (i.e., any attempt by Eve to cause $s_3^\prime \neq s_3$), and that the verification process will also not reveal any information about $s_3$ to Eve.

We note that any modification by Eve of the messages exchanged by Alice and Bob during the verification process is equivalent to Eve introducing errors in $s_1^\prime$ and $s_2^\prime$ during the main part of the relay protocol. If she controls at least one intermediate node, Eve can introduce such errors by modifying one or more of the strings transmitted by a node under her control. We can thus completely describe Eve's attack on the protocol by a string $e=(e_1,e_2,e_3)$, where $s^\prime = s \oplus e$, and the three substrings $e_1$, $e_2$, and $e_3$ have lengths $\ell_1$, $\ell_2$, and $\ell_3$, respectively (with $\ell = \ell_1+\ell_2+\ell_3$).

It is clear that Eve cannot gain any information about $s_3$ from the verification process, since the only information ever transmitted about $s_3$ (the hash $H[s_3]$) is encrypted by the OTP $s_1$, and $s_1$ is never re-used.

Before proceeding, let us further partition $s_1$ into two strings $s_{1a}$ and $s_{1b}$, where $s_{1a}$ is the portion of $s_1$ used to encrypt $r$, and $s_{1b}$ is the portion used to encrypt $H[s_3]$.  Let $\ell_{1a}$ and $\ell_{1b}$ be the lengths of $s_{1a}$ and $s_{1b}$. We similarly partition $s_1^\prime$ and $e_1$.

Eve's only hope of fooling Bob into accepting a tampered-with key (i.e., accepting even though $s_3^\prime \neq s_3$) is for her to choose $e_{1b}$ and $e_3$ such that the expression $H[s_3]\oplus H[s_3 \oplus e_3] =  e_{1b}$ is satisfied.  Random guessing will give her a $\sim2^{-\ell_{1b}}$ chance of tricking Bob into accepting; for Eve to do better, she must be able to exploit a weakness in the hash function $H$ that gives her some information as to the correct value of $e_{1b}$ for some choice of $e_3$. Note that Eve's best strategy for this attack is to choose $e_{1a}$ and $e_2$ to be just strings of zeroes.

From this observation, we obtain the following condition on the hash function: for a random $s_3$ (unknown to Eve), there exists no choice of $e_3$ such that Eve has any information about the value of $e_{1b}$ she should choose to satisfy $H[s_3]\oplus H[s_3 \oplus e_3] =  e_{1b}$. In practice, it would be acceptable for Eve to gain a very small amount of information, as long as the information gained did not raise Eve's chances much beyond random guessing. This is a relatively weak requirement on $H$, and is likely satisfied by any reasonable choice of hash function.

To fool Alice into falsely accepting, Eve can either fool Bob via the aforementioned method, or Eve can attempt to impersonate Bob by sending Alice a random string of length $\ell_2$, in the hopes that it happens to be equal to $s_2 \oplus H[r]$. Clearly, her chances for the latter method are no better than $2^{-\ell_2}$. The latter method of attack only fools Alice and not Bob; it is thus of limited use to Eve.

We note that the security of the verification protocol depends on the choice of $\ell_1$ and $\ell_2$ (as described above); these parameters should be chosen so as to provide whatever degree of security is required. Alice and Bob choose $\ell_3$ so as to obtain whatever size key they desire. Since the security of the verification process does not depend on $\ell_3$, the communication cost of key verification is negligible in the limit of large $\ell_3$ (i.e., in the limit of large final key size).

\section{Security of the Relay Protocol\label{sec:security}}
In order for the secret to be compromised, there must be some $j \in \{1, \ldots, m-1\}$ such that, for all $i \in \{1, \ldots, n\}$,  at least one of $v_{i,j}$ and $v_{i,j+1}$ is dishonest (i.e., such that, for some $j$, every string $r_{i,j}$ is either sent or received by a compromised party). If this happens, we say the protocol has been compromised at stage $j$.  For a given $j$, the probability of compromise is $(1-t^2)^n$, but the probability for $j$ is not entirely independent of the probabilities for $j-1$ and $j+1$. Thus, we can bound from below the overall probability of the channel between Alice and Bob being secure, $p_s$, by (\ref{eq:relay_bounds}):
\begin{eqnarray}
p_s & \geq & \left[1- (1-t^2)^n\right]^{m-1}. \label{eq:relay_bounds}
\end{eqnarray}
From this result, we see that, if we wish to ensure our probability of a secure channel between Alice and Bob is at least $p_s$, it is sufficient to choose $n = \log \left( 1- p_s^{1/(m-1)} \right) / \log \left(  1- t^2 \right)$. Intercity bandwidth consumed is proportional to $n$, so we see that we have good scaling of resource consumption with communication distance. Alternatively, we can re-write the equation for choosing $n$ in terms of a maximum allowed probability of compromise, $\delta = 1 - p_s$. For $\delta \ll 1$, we obtain the following relation:
\begin{eqnarray*}
n & \simeq & \frac{\log{(m-1)} - \log {\delta}}{-\log {(1 -t^2)}}.
\end{eqnarray*}
Total resource usage (intercity communication links required) scales as $\mathcal{O}(mn)$, or $\mathcal{O}(m \log{m})$ for fixed $\delta$, $t$. While intracity communication requirements scale faster (as $\mathcal{O}(mn^2)$), it is reasonable to ignore this because of the comparatively low cost of intracity communication and the finite size of the earth (which effectively limits $m$ to a maximum of 100 or so for a QKD network with single link distances of $\sim100\ \mathrm{km}$).

If each party in the network simultaneously wished to communicate with one other party (with that party assumed to be $m/2$ cities away on average), total intercity bandwidth would scale as $\mathcal{O}(m^2n^2)$. By comparison, the bandwidth for a network of the same number of parties employing public key cryptography (and no secret sharing) would scale as $\mathcal{O}(m^2n)$. Since $n$ scales relatively slowly (i.e., with $\log m$), this is a reasonable penalty to pay for improved security.

\section{Alternative Adversary Models}
We now briefly consider a number of alternative adversary models. First, let us consider replacing adversary capability (\ref{it:adcap3}) with the following alternative, which we term (\ref{it:adcap3}$^\prime$): the adversary can compromise up to $k-1$ nodes of its choice. Compromised nodes are assumed to be under the complete control of the adversary, as before. In this scenario, the security analysis is trivial. If $k > n$, the adversary can compromise Alice and Bob's communications undetected. Otherwise, Alice and Bob can communicate securely. 

We could also imagine an adversary controls some random subset of nodes in the network---as described by (\ref{it:adcap3})---and wishes to disrupt communications between Alice and Bob (i.e., perform a DOS attack), but does not have the capability to disrupt or modify public channels. Alice and Bob can modify the protocol to simultaneously protect against both this type of attack and also the adversary mentioned in Section \ref{sec:ad_cap}. To do so, they replace the simple secret sharing scheme described above with a Proactive Verifiable Secret Sharing (PVSS) scheme~\cite{darco:vss}. In this scenario, nodes can check at each stage to see if any shares have been corrupted, and take corrective measures. This process is robust against up to $n/4 - 1$ corrupt shares, which implies that PVSS yields little protection against DOS attacks unless $t > t_{\mathrm{thresh}} \approx \sqrt{3}/2$.

\section{Conclusion\label{sec:conclusion}}

We have shown a protocol for solving the relay problem and building secure long-distance communication networks with present-day QKD technology. The protocol proposed employs secret sharing and multiple paths through a network of partially-trusted nodes. Through the choice of moderately large $n$ in the relay problem, one can make the possibility of compromise vanishingly small. For fixed probability of compromise of each of the intermediate nodes, the number of nodes per stage required to maintain security scales only logarithmically with the number of stages (i.e., with distance).

Given that QKD systems are already commercially available, our methods could be implemented today.

\section{Acknowledgments}
We wish to thank Louis Salvail, Aidan Roy, Rei Safavi-Naini, Douglas Stebila, Hugh Williams, Kevin Hynes, and Renate Scheidler for valuable discussions. TRB acknowledges support from a US Department of Defense NDSEG Fellowship. BCS acknowledges support from iCORE and CIFAR.

\bibliographystyle{splncs}
\bibliography{authentication}

\begin{thebibliography}{10}

\bibitem{shor:factor2}
Shor, P.W.:
\newblock {Algorithms for quantum computation: Discrete logarithms and
  factoring}.
\newblock Proc.\ of 35th Annual Symposium on Foundations of Computer Science
  (1994)  124--134

\bibitem{bennett:BB84}
Bennett, C.H., Brassard, G.:
\newblock Quantum cryptography: Public key distribution and coin tossing.
\newblock Proc.\ of IEEE International Conference on Computers, Systems, and
  Signal Processing (1984)  175--179

\bibitem{takesue:qkd}
Takesue, H., Nam, S.W., Zhang, Q., Hadfield, R.H., Honjo, T., Tamaki, K.,
  Yamamoto, Y.:
\newblock Quantum key distribution over a 40-db channel loss using
  superconducting single-photon detectors.
\newblock Nature Photonics \textbf{1} (2007)  343--348

\bibitem{wootters:cloning}
Wootters, W.K., Zurek, W.H.:
\newblock A single quantum cannot be cloned.
\newblock Nature \textbf{299} (1982)  802--803

\bibitem{briegel:5932}
Briegel, H.J., Dur, W., Cirac, J.I., Zoller, P.:
\newblock Quantum repeaters: The role of imperfect local operations in quantum
  communication.
\newblock Phys.\ Rev.\ Lett. \textbf{81} (1998)  5932--5935

\bibitem{duan:6862}
Duan, L.M., Lukin, M., Cirac, J.I., Zoller, P.:
\newblock Long-distance quantum communication with atomic ensembles and linear
  optics.
\newblock Nature \textbf{414} (2001)  413--418

\bibitem{ekert:661}
Ekert, A.K.:
\newblock {Quantum cryptography based on Bell's theorem}.
\newblock Phys.\ Rev.\ Lett. \textbf{67} (1991)  661--663

\bibitem{simon:190503}
Simon, C., de~Riedmatten, H., Afzelius, M., Sangouard, N., Zbinden, H., Gisin,
  N.:
\newblock Quantum repeaters with photon pair sources and multimode memories.
\newblock Phys.\ Rev.\ Lett. \textbf{98} (2007)  190503

\bibitem{shamir:secret}
Shamir, A.:
\newblock How to share a secret.
\newblock Comm.\ of the ACM \textbf{22} (1979)  612--613

\bibitem{blakley:313}
Blakley, G.R.:
\newblock Safeguarding cryptographic keys.
\newblock Proc.\ of the National Computer Conference \textbf{48} (1979)
  313--317

\bibitem{renner:012332}
Renner, R., Gisin, N., Kraus, B.:
\newblock Information-theoretic security proof for quantum-key-distribution
  protocols.
\newblock Physical Review A \textbf{72} (2005)  012332

\bibitem{ben-or:distributed}
Ben-Or, M., Goldwasser, S., Wigderson, A.:
\newblock Completeness theorems for non-cryptographic fault-tolerant
  distributed computation.
\newblock Proc.\ of the 20th Annual ACM Symposium on Theory of Computing (1988)
   1--10

\bibitem{ostrovsky:112605}
Ostrovsky, R., Yung, M.:
\newblock How to withstand mobile virus attacks.
\newblock Proc.\ of the 10th Annual ACM Symposium on Principles of Distributed
  Computing (1991)  51--59

\bibitem{herzberg:339}
Herzberg, A., Jarecki, S., Krawczyk, H., Yung, M.:
\newblock Proactive secret sharing, or how to cope with perpetual leakage.
\newblock In: Advances in Cryptology -- CRYPTO 1995. Volume 963 of Lecture
  Notes in Computer Science. (1995)  339--352

\bibitem{salvail:qkd}
Salvail, L.:
\newblock {Security Architecture for SECOQC: Secret-key Privacy and
  Authenticity over QKD Networks}.
\newblock {D-SEC-48}, SECOQC (2007)

\bibitem{darco:vss}
D'Arco, P., Stinson, D.R.:
\newblock On unconditionally secure robust distributed key distribution
  centers.
\newblock In: Advances in Cryptology -- ASIACRYPT 2000. Volume 2501 of Lecture
  Notes in Computer Science. (2002)  346--363

\end{thebibliography}

\end{document}